\documentclass[reprint,showpacs,floatfix,amsmath,amssymb,prb]{revtex4-2}
\usepackage{graphicx,tikz,latexsym,hyperref,colortbl}
\usepackage{multirow}
\usepgflibrary{shapes.arrows}
\usepgfmodule{decorations} \usepackage{fancyhdr} \usepackage{textcomp}

\newcommand{\fref}[1]{Figure~\ref{fig:#1}}
\newcommand{\tref}[1]{Table~\ref{tab:#1}}

\hypersetup{colorlinks=false, citecolor=purple, linkcolor=red}

\begin{document}
\title{  Canted antiferromagnetism in high purity $\mathrm{NaFeF_3}$ prepared by a novel wet-chemical synthesis method }

\author{Fabian L. M. Bernal}
\affiliation{Chemistry Department and Center for Material Science and Nanotechnology, University of Oslo, NO-0315, Oslo, Norway}
\author{Bruno Gonano}
\affiliation{Chemistry Department and Center for Material Science and Nanotechnology, University of Oslo, NO-0315, Oslo, Norway}
\author{Fredrik Lundvall}
\affiliation{Chemistry Department and Center for Material Science and Nanotechnology, University of Oslo, NO-0315, Oslo, Norway}
\author{David S. Wragg}
\affiliation{Chemistry Department and Center for Material Science and Nanotechnology, University of Oslo, NO-0315, Oslo, Norway}
\author{Helmer Fjellv\aa g}
\affiliation{Chemistry Department and Center for Material Science and Nanotechnology, University of Oslo, NO-0315, Oslo, Norway}
\author{Fabien Veillon}
\affiliation{Laboratory Crismat, UMR6508 CNRS, Normandie University, ENSICAEN, UNICAEN, 6 bd Mar\'echal Juin, 1450 Caen cedex 4, France}
\author{Wojciech A. S\l awi\'{n}ski}
\affiliation{Faculty of Chemistry, University of Warsaw, Pasteura 1, 02-093 Warsaw, Poland}
\affiliation{ISIS Facility, Rutherford Appleton Laboratory, Harwell Oxford, Didcot, Oxfordshire OX11 0QX, U.K.}
\author{\O ystein S. Fjellv\aa g}
\affiliation{Department for Neutron Materials Characterization, Institute for Energy Technology, PO Box 40, NO-2027, Kjeller, Norway}

\begin{abstract}
  We report a novel synthesis  method for, and structural and magnetic
  characterization of the fluoroperovskite $\mathrm{NaFeF_3}$. We have
  developed a  wet-chemical method  that allows preparation  of large
  volumes  of   air-sensitive  fluoroperovskites  with   high  purity.
  $\mathrm{NaFeF_3}$ has  a N\'eel temperature  ($T_N$) of 90 K  and a
  Weiss  constant  ($\theta$) of  -124  K,  corresponding to  dominant
  antiferromagnetic interactions.  Below $T_N$, a slight difference is
  observed  between zero-field  and field  cooled samples,  indicating
  spin-canting and weak ferromagnetism.  AC magnetometry confirms that
  weak ferromagnetism is inherent to $\mathrm{NaFeF_3}$ and not due to
  impurities. From  powder neutron  diffraction data, we  describe the
  magnetic  structure precisely  as a  weakly canted  G-type (magnetic
  space  group $Pn'ma'$).   A  ferromagnetic component  is allowed  in
  $Pn'ma'$, however,  this component  may be  absent in  zero magnetic
  fields and is  too small  to  be confirmed  on the  basis of  powder
  neutron diffraction data.
\end{abstract}

\maketitle

\section{Introduction}
\label{sec:intro}

Fluoroperovskites  display rich  structural chemistry,  strongly ionic
bonding character (due  to the high electronegativity  of the fluoride
anions), and  corresponding localized electron  magnetism \cite{FLUOR,
  FLUOR2}.  They  exhibit a  wide  range  of  properties that  can  be
utilized  in  e.g.  data storage,  computer  processors,  spintronics,
multiferroics  and  batteries   \cite{SPINTRON1,  SPINTRON2,  MultiF1,
  MultiF2}.

${\rm NaFeF_3}$ has attracted attention as a low-cost cathode material
\cite{NF3SIB,NF3SIB2}. Its advantages in  this application include the
Earth's  abundance  of  the   constituent  elements,  intrinsic  anion
stability and  a theoretical capacity  of 197 ${\rm mAhg^{-1}}$  for a
one-electron process.  ${\rm NaFeF_3}$  nanoplates in  particular show
good capacity retention compared to other metal fluoride and composite
cathode materials with 50\% retained  capacity for Na after 200 cycles
at 0.2 A g$^{-1}$ \cite{NCATH}.

The ${\rm  NaFeF_3}$ fluoroperovskite  has intriguing  phase relations
under  high-pressure.   At  ambient   temperature  and  pressure,  the
compound  adopts  the  orthorhombic  perovskite  ${\rm  GdFeO_3}$-type
crystal  structure  with space  group  $Pnma$.  It transforms  into  a
corrugated   layered   ${\rm  CaIrO_3}$-type   post-perovskite   (pPv)
structure  at  room  temperature  at   9  GPa  \cite{ME1}.   A  second
structural phase transition occurs at 20 GPa from pPv-to-ppPv with the
${\rm  Sb_2S_3}$-type crystal  structure  \cite{ME2}.  The  remarkable
structural flexibility of  ${\rm NaFeF_3}$ has made  it an interesting
candidate  for  studies  and   simulations  of  extreme  environments,
including the Earth's interior and exoplanets \cite{PPV}.

The  electronic   configuration  of   the  ${\rm  Fe^{2+}}$   ions  in
${\rm NaFeF_3}$  is high-spin (HS)  $t_{2g}^4 e_g^2$. They  follow the
spin-only model  with $S=2$ and  a theoretical paramagnetic  moment of
$\mu_{eff}=4.90 \mu_B$,  which is influenced by  orbital contributions
and usually leads to slightly higher $\mu_{eff}$.

${\rm   Fe^{2+}}$   is   air   sensitive  and   oxidizes   easily   to
${\rm   Fe^{3+}}$.     Controlling   the   anaerobic    chemistry   of
${\rm Fe^{2+}}$ ions  is a prerequisite for synthesis  of single phase
${\rm NaFeF_3}$. We  have previously utilized a  solid state synthesis
under inert conditions \cite{ME1}, yielding a sample with $<0.5$ \% Fe
metal  impurity. Impurities,  either  from an  incomplete solid  state
reaction  or  from  the  iron reactor,  may  introduce  inaccuracy  to
magnetometric studies and disguise  the intrinsic magnetic behavior of
${\rm NaFeF_3}$.   Indeed, to the best  of our knowledge there  are no
neutron powder diffraction  (NPD) studies on the  magnetic ordering in
${\rm   NaFeF_3}$,   probably   because   the   air   sensitivity   of
${\rm  Fe^{2+}}$ makes  it very  difficult to  produce sample  volumes
sufficient for NPD experiments by conventional solid state methods.

In  this article  we  describe a  wet-chemistry  synthesis method  for
${\rm  NaFeF_3}$, which  produces  iron-free  material in  substantial
volumes. Using the  high purity ${\rm NaFeF_3}$ samples,  we study the
intrinsic magnetic  properties of the compound.  Further, the magnetic
structure is  precisely described based on  neutron powder diffraction
data.

\section{Experimental}
\label{sec:exp}
\subsection{Synthesis of $\mathrm{NaFeF_3}$}
\label{sec:syn}
${\rm  NaFeF_3}$  was  synthesised  on a  Schlenk-line  equipped  with
flexible  hoses.  Two  polycarbonate vials  of  85 and  200 ml  volume
(denoted A and B respectively) were  used for the reaction. Vial A was
filled with 2  g of Fe ($\sim$0.035 g mol$^{-1}$,  99.999 \% pure) and
vial  B with  0.08 mol  ($\sim$ 3.35  g) NaF.   The vials  were closed
tightly with silicone rubber septa, connected through the hoses to the
Schlenk-line  and  thoroughly  flushed  with   Ar.  The  Ar  flow  was
maintained  throughout  the reaction  to  ensure  inert conditions.  A
needle  was placed  in each  septum to  vent the  excess gas  from the
vials. 10  mL of  HCl (37 \%)  and 20 mL  ${\rm H_2O}$  were degassed,
mixed  and  added to  vial  A.  20 mL  of  degassed  ${\rm H_2O}$  was
carefully injected (using  a syringe first evacuated  and flushed with
Ar) to the NaF in vial B.  Both vials were placed in an oil-bath under
constant Ar  flow at 90 $^\circ$C  until the oxidation of  Fe metal to
${\rm FeCl_2}$ was completed.  The ${\rm FeCl_2}$ solution from vial A
was then quickly transferred to vial B with an Ar flushed syringe. The
contents of vial B were stirred constantly during injection to mix the
two solutions. Thereafter,  vial B was cooled to 80  $^\circ$C and the
contents stirred for  30 to 60 minutes. ${\rm NaFeF_3}$  appeared as a
beige  precipitate. The  product was  washed repeatedly  with degassed
water and MeOH under flowing Ar,  with decanting the liquid after each
washing.   Finally  the  solid  product was  removed  from  the  vial,
filtered, washed thoroughly with degassed MeOH, and dried under vacuum
overnight before  storing in the  Ar-atmosphere of a glove  box. Phase
purity  was   confirmed  by   powder  x-ray  diffraction   (PXRD)  and
magnetometry.

\subsection{Powder X-ray diffraction}
\label{sec:sc}

PXRD  data  for $\mathrm{NaFeF_3}$  were  collected  at the  Norwegian
National Resource Centre for X-ray Diffraction, Scattering and Imaging
(RECX) on a  Bruker D8 Advance diffractometer in  capillary mode using
$\mathrm{Cu_{K\alpha 1}}$ radiation (1.540598Å)  selected by a Ge(111)
focusing monochromator, and a LynxEye XE detector system.

\subsection{Magnetic characterization}
\label{sec:mc}

Magnetometry experiments  were performed  with a 9T  Physical Property
Measurement System (PPMS, Quantum Design)  on 48 mg of polycrystalline
powder. Temperature  dependent DC  magnetic susceptibility  $\chi (T)$
measurements were conducted between 2 and  300 K for zero field cooled
samples, followed by  studies at field cooled conditions  (ZFC and FC,
respectively).   The   magnetic   susceptibility  is   calculated   by
$\chi   =M/H$    where   $M$   is   the    magnetization,   given   in
emu.$\mathrm{mol^{-1}  Oe^{-1}}$   and  H   the  magnetic   field  (10
kOe). Isothermal  field dependent measurements $M  (H)$ were collected
at  ~2K, likewise  half-loop isothermal  measurements at  70, 120  and
300~K,  all up  to  90  kOe. AC  measurements  were  carried out  with
frequencies ranging from 100 Hz to 10 kHz with a 10 Oe field.

\subsection{Neutron Powder Diffraction}
\label{sec:npd}

NPD data for $\mathrm{NaFeF_3}$ was measured on the WISH instrument at
the ISIS pulsed neutron and muon source (UK) \cite{WISH}.  Diffraction
patterns were collected between 2 and  297 K, and the data was reduced
with the  Mantid software \cite{MANTID}.   Data from the  four highest
resolution detector banks were used, as the lowest resolution bank did
not contain  any unique  information.  NPD  was collected  at selected
temperatures below and above $T_N$.

The magnetic refinements were carried  out in the magnetic space group
$Pn'ma'$, which allows non-zero values  for $M_x$, $M_y$ and $M_z$, in
the  Jana2006 software  \cite{JANA}. The  background (5  term Legendre
polynomials),  peak-shape,  atomic  positions (according  to  symmetry
restrictions),  isotropic  thermal  displacement parameters  for  each
element type,  lattice parameters  and scale parameters  were refined.
The derived values for $M_x$ and $M_z$ are given in \tref{moment}.

\section{Results}
\label{sec:RES}

\subsection{Synthesis procedure and crystal structure}

Fluoroperovskites are typically prepared  by solid-state reactions and
may  contain  magnetic  impurities  that  make  them  appear  as  weak
ferromagnets,  both  below  and   above  the  N\'eel  temperature,  as
e.g.    $\mathrm{\alpha}$-Fe    impurities    in    $\mathrm{NaFeF_3}$
\cite{ME1}.  The  air-sensitive   chemistry  of  fluoroperovskites  is
demanding.  We currently  benefit from a new  wet-chemical method that
bypass the challenges of  conventional solid-state reactions by always
working under inert conditions on  a Schlenk line.  This wet-chemistry
approach  is  ideal  for  synthesis  of  air-sensitive  fluorides,  as
recently  shown  for  the   extremely  air-sensitive  ${\rm  Cr^{2+}}$
\cite{ME3}.  $\mathrm{NaFeF_3}$ is currently prepared by means of this
wet-chemical method, and high purity was confirmed by XRD and NPD.

$\mathrm{NaFeF_3}$      adopts     the      distorted     orthorhombic
$\mathrm{GdFeO_3}$ perovskite  structure with  space group  $Pnma$ and
Glazer  tilt $\mathrm{a^-b^+a^-}$,  \fref{str}.  The  relation to  the
ideal        cubic        perovskite         is        given        by
$  a \approx  c \approx  \sqrt{2}a_{c}$ and  $b \approx  2a_{c}$ where
$a_{c}$  is the  lattice parameter  of  the cubic  perovskite. A  weak
Jahn-Teller distortion is  present in the system  originating from the
high-spin  $d^6$  electron  configuration of  $\mathrm{Fe^{2+}}$.  The
consequence of  the weak Jahn-Teller  effect is slight  differences in
the bond  lengths; the  Fe-F1 bonds adopt  a medium  length (2.0744(3)
\AA), while Fe-F2 forms two short  and two long bonds (2.0564(6)~\AA ~
and 2.0795(6)~\AA~  respectively), \fref{str}. Bond length  values are
from NPD at 297 K, see below.

\begin{figure*}[ht!]
  \centering
  \includegraphics[scale=.24]{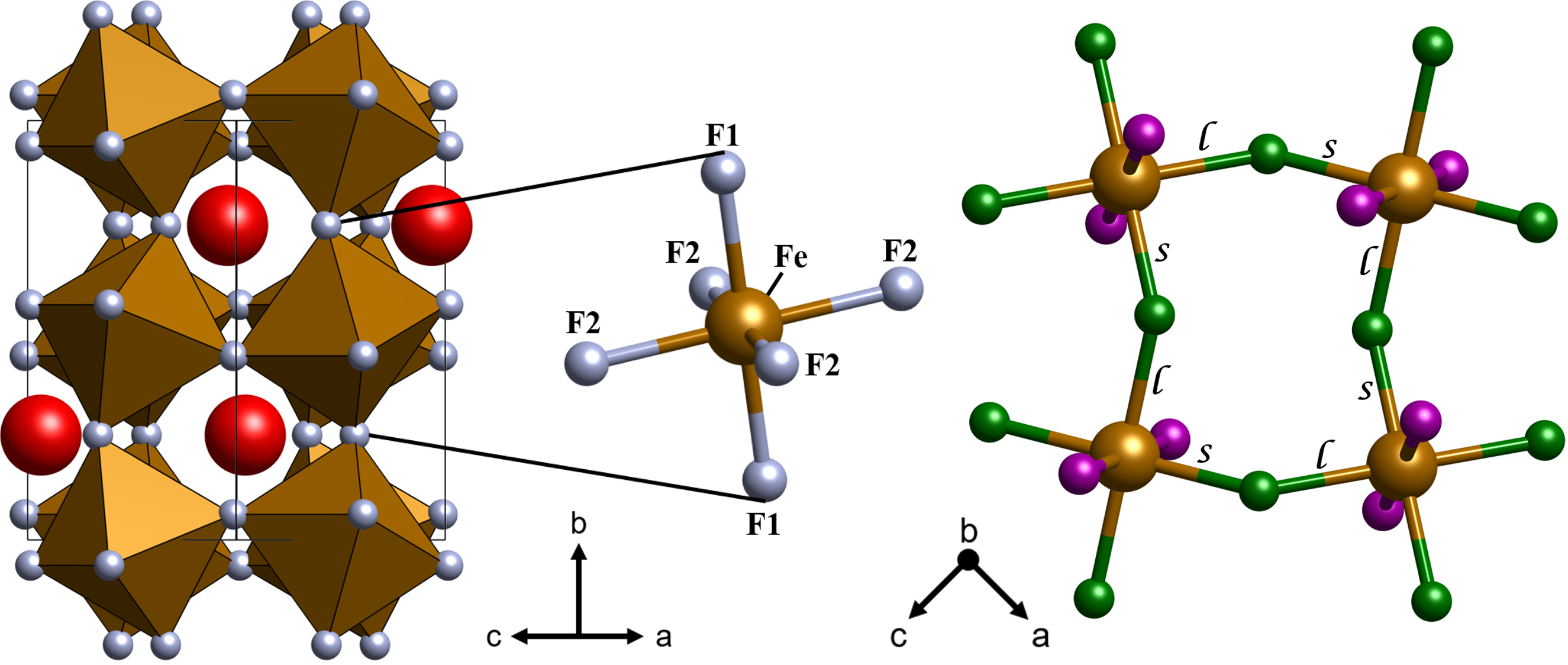}
  \caption{Crystal   structure   of    the   orthorhombic   perovskite
    $\mathrm{NaFeF_3}$   ($Pnma$).    Left:   Illustration    of   the
    corner-connected  octahedra of  $\mathrm{FeF_6}$ with  Glazer tilt
    $\mathrm{a^-b^+a^-}$.  Sodium,  iron and fluorine atoms  are shown
    in red, orange and gray,  respectively.  Right: Illustration of the
    $ac$-plane, where ordered short and long Fe-F2 bonds are formed by
    the weak  Jahn-Teller effect. The  short and long Fe-F2  bonds are
    marked  $s$ and  $l$ respectively,  while medium  Fe-F1 bonds  are
    out-of-plane.  Fe, F1 and F2 are shown in orange, purple and green,
    respectively.  }
  \label{fig:str}   
\end{figure*}

\subsection{Magnetic properties}
Variable temperature DC magnetization measurements were carried out on
a polycrystalline  sample between  2 and  300 K under  a 10  kOe field
(\fref{mag}).    The    data    are   consistent    with    long-range
antiferromagnetic  ordering. A  sharp decrease  in the  molar magnetic
susceptibility is associated  with a N\'eel transition  at 90~K.  From
the  inverse  susceptibility  $\chi^{-1}$ curve  in  the  paramagnetic
region  (100  to  300~K),  we   calculate  a  paramagnetic  moment  of
$\mu_{eff} =  5.58~\mu_B$. This  is in  good agreement  with typically
observed  values  for  ${\rm Fe^{2+}}$  (5.0-5.6~$\mu_B$).   From  the
Curie-Weiss fit, we  extract a Weiss-temperature of  $\theta$ = -124~K
(data  measured  in a  field  of  10~kOe), confirming  the  dominating
antiferromagnetic nature of ${\rm NaFeF_3}$.

\begin{figure}[ht!]
  \centering
  \includegraphics[scale=.30]{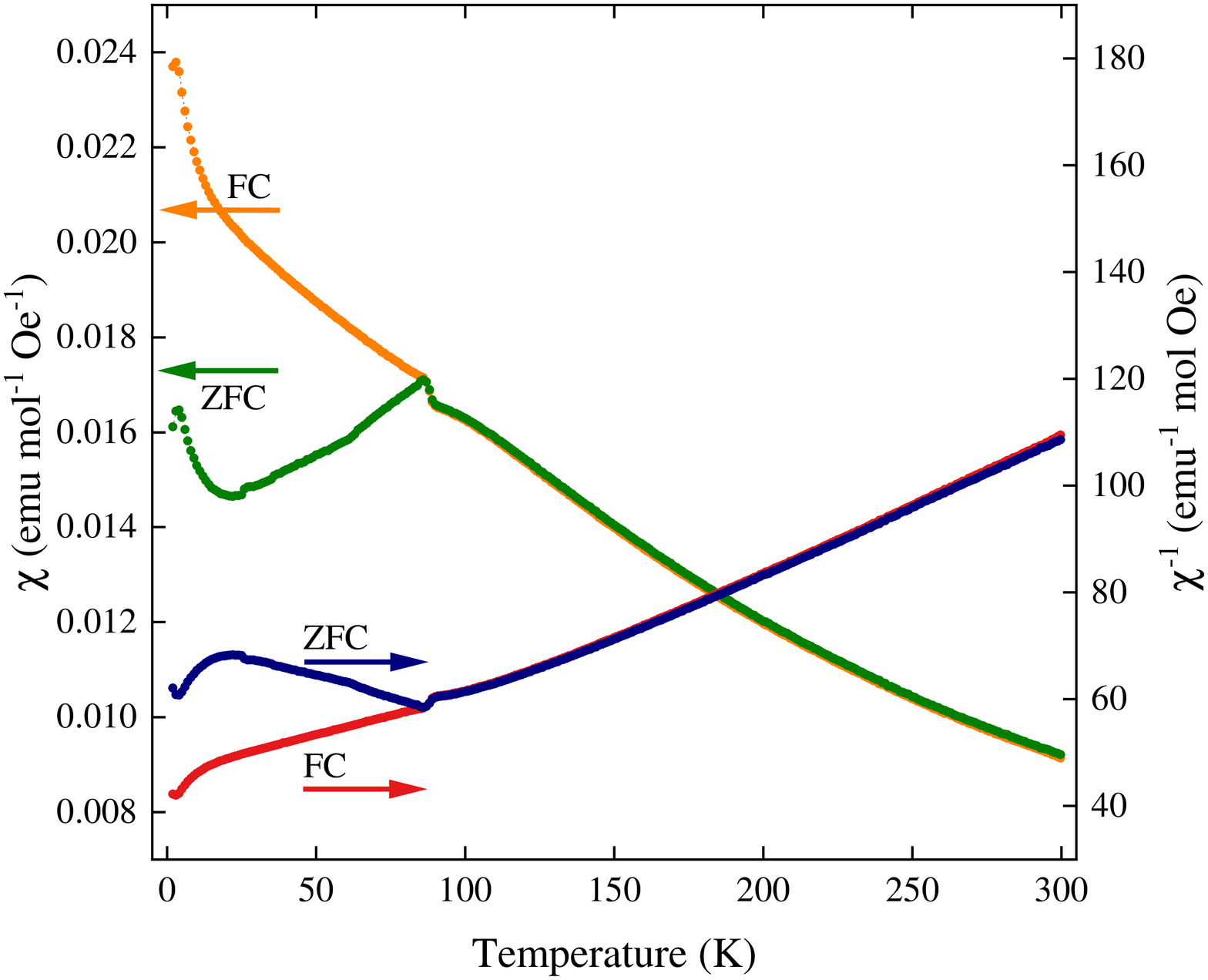}
  \caption{Temperature  dependency  of   the  magnetic  susceptibility
    $\chi(T)$ at $H$ = 10 kOe  measured in ZFC-FC mode (left axis) and
    inverse $\chi^{-1}$ (right axis).}
  \label{fig:mag}
\end{figure}

We note a significant difference between  the FC and ZFC curves at low
temperatures, as  well as a  minor hysteresis.  This might  indicate a
transition to a spin-glass like state at low temperatures. However, AC
magnetization  measurements  (\fref{X}a) in  a  10  Oe field  show  no
variations of  the N\'eel temperature  versus frequency for  $\chi '$,
refuting this hypothesis.

\begin{figure}[ht!]
  \centering
  \includegraphics[scale=.30]{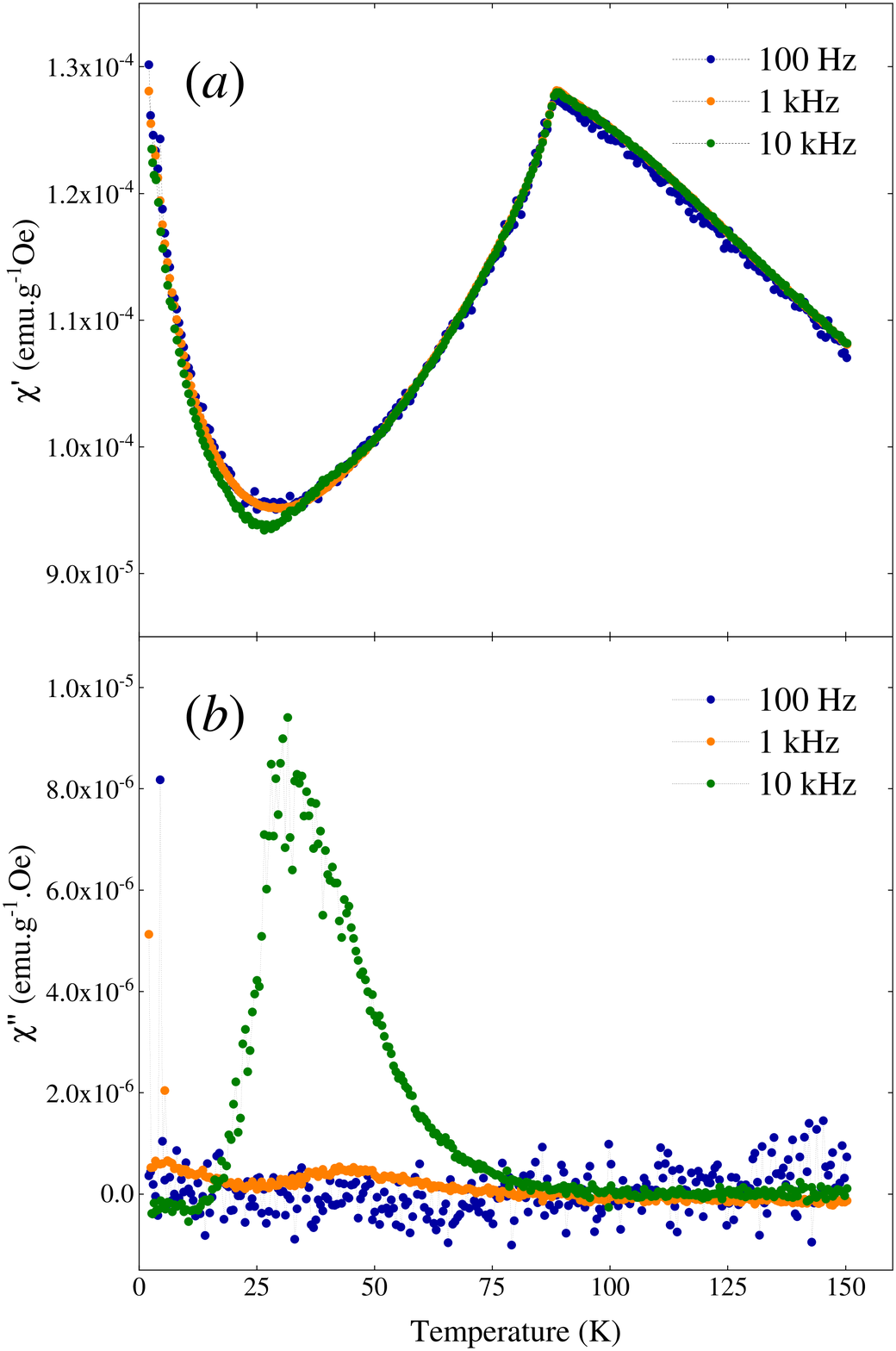}
  \caption{ Temperature dependence of the (\emph{a}) real ($\chi'$) and (\emph{b}) imaginary ($\chi''$) part of the AC magnetic
    susceptibility   at different  frequencies (100  Hz, 1 kHz
    and  10 kHz).  
  }
  \label{fig:X}
\end{figure}

Appearance of magnetic hysteresis may  be explained by the presence of
a small ferromagnetic moment  originating from spin canting, resulting
in weak ferromagnetism. For a purely antiferromagnetic transition, the
imaginary  component  $\chi   ''$  is  expected  to  be   zero  in  AC
measurements  \cite{ACmag}.  In fact,  for  $\chi  ''$ (\fref{X}b)  we
observe a  strong peak for  an AC  field of 10  kHz. The peak  is less
pronounced for 1 kHz AC field, while only noise is observed for 100 Hz
AC  field. The  peak starts  to increase  in intensity  at the  N\'eel
temperature,  indicating  that  it  is associated  with  the  magnetic
transition  in $\mathrm{NaFeF_3}$,  and that  we can  excluded effects
caused  by  an impurity,  such  as  e.g. $\mathrm{\alpha}$-Fe.  It  is
therefore evident that spin-canting  results in weak ferromagnetism in
$\mathrm{NaFeF_3}$,  inherent  to  the  compound.  We  note  that  the
ferromagnetism  is very  weak  and  based on  DC  (\fref{mag}) and  AC
(\fref{X}) magnetometry, it should be very close to zero in the absent
of  a magnetic  field. The  maximum of  the peak  is at  $\sim$ 32  K,
indicating  that  the spin-canting  is  further  developing below  the
N\'eel temperature.

Isothermal  field dependent  magnetic  measurements  above the  N\'eel
temperature ($T_N$ =  90~K) at 120 and 300~K show  a linear behaviour,
associated   with  a   paramagnetic  state   (\fref{hyst}).  The   low
magnetization observed at 90 kOe ($0.25~\mu_B$/Fe at 2~K) confirms the
dominating antiferromagnetic  behavior of the system.   However, below
the  N\'eel   temperature,  a  slight  hysteresis   is  observed.  The
hysteresis is  clear at 2~K (inset  in \fref{hyst}), while it  is less
prominent at  70~K. The  presence of  hysteresis below  $T_N$ supports
that weak ferromagnetism is intrinsic to the compound.

\begin{figure}[t!]
  \centering
  \includegraphics[scale=.30]{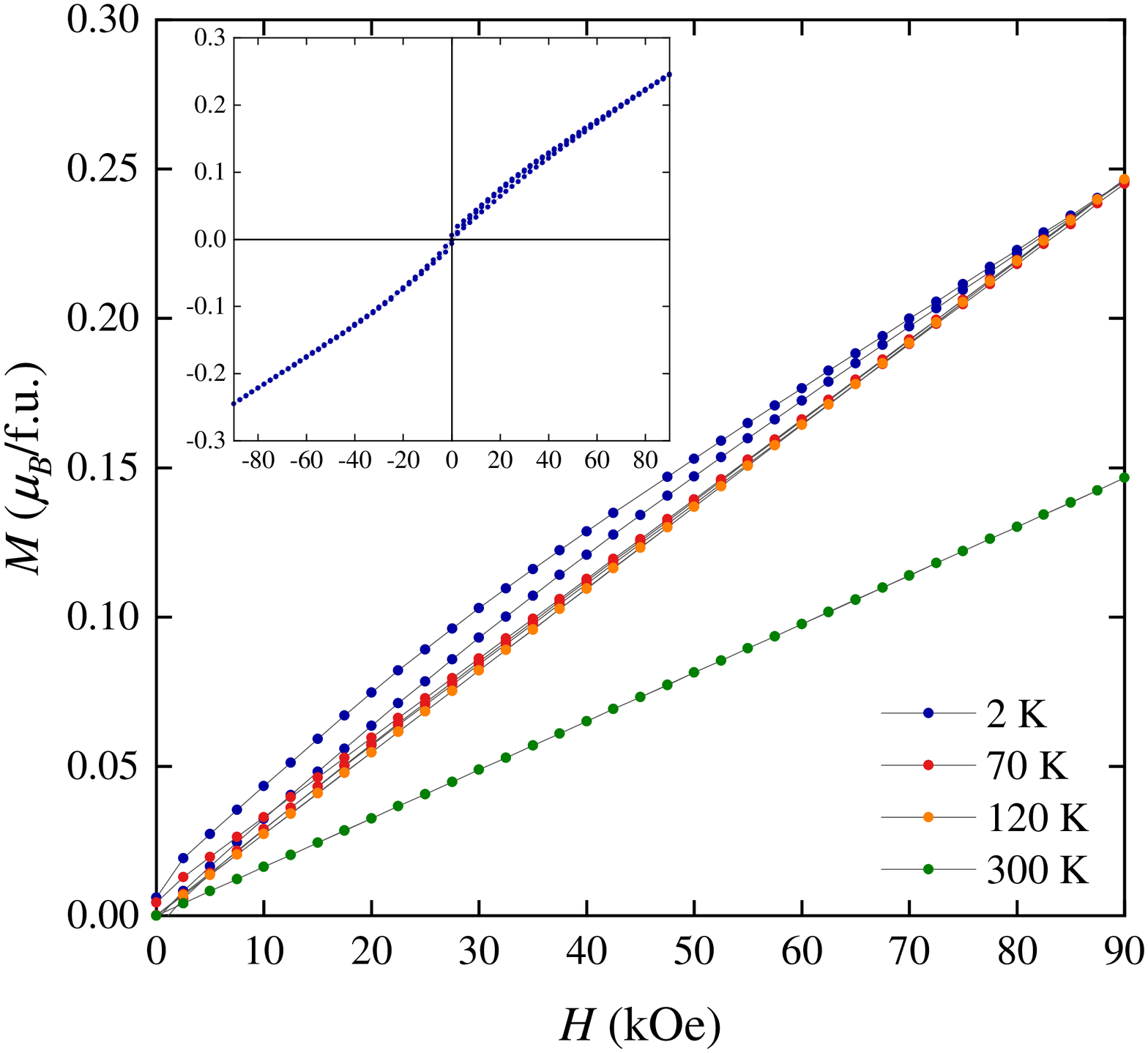}
  \caption{Isothermal $M$ versus $H$ curves  with an applied field 0~T
    $\to$ 9~T  $\to$ 0~T  recorded at  2, 70,  120 and  300~K. Insert:
    $M$ versus $H$ curve at  2 K to  show the full symmetry.}
  \label{fig:hyst}
\end{figure}

\subsection{Neutron diffraction and magnetic structure}
Neutron diffraction was carried out between 2 and 297~K to investigate
possible ordering of ${\rm Fe^{2+}}$ magnetic moments.  At 2~K, strong
additional reflections from long-range magnetic ordering are evident ,
e.g. two  strong reflections  at 4.49  and 4.58 \AA  ~ and  one weaker
reflection  at  7.89  \AA   ~(\fref{NPD}  and  \fref{NPD_RT}).   These
reflections were  indexed according  to the unit  cell of  the crystal
structure of  $\mathrm{NaFeF_3}$, and corresponds to  a $\Gamma$-point
magnetic propagation vector  $k = (0, 0, 0)$. However,  several of the
observed magnetic reflections break  the symmetry extinctions of space
group $Pnma$. We hence evaluated  the possible magnetic structures for
the   $\Gamma$-point   representations  (\tref{irrep})   by   Rietveld
refinements  against NPD  at  2~K,  and we  found  $  \Gamma^+_4 $  to
precisely  describe the  magnetic  ordering. This  corresponds to  the
magnetic space group $Pn'ma'$ (\fref{magstr}).

\begin{figure}[ht!]
  \centering
  \includegraphics[scale=.34]{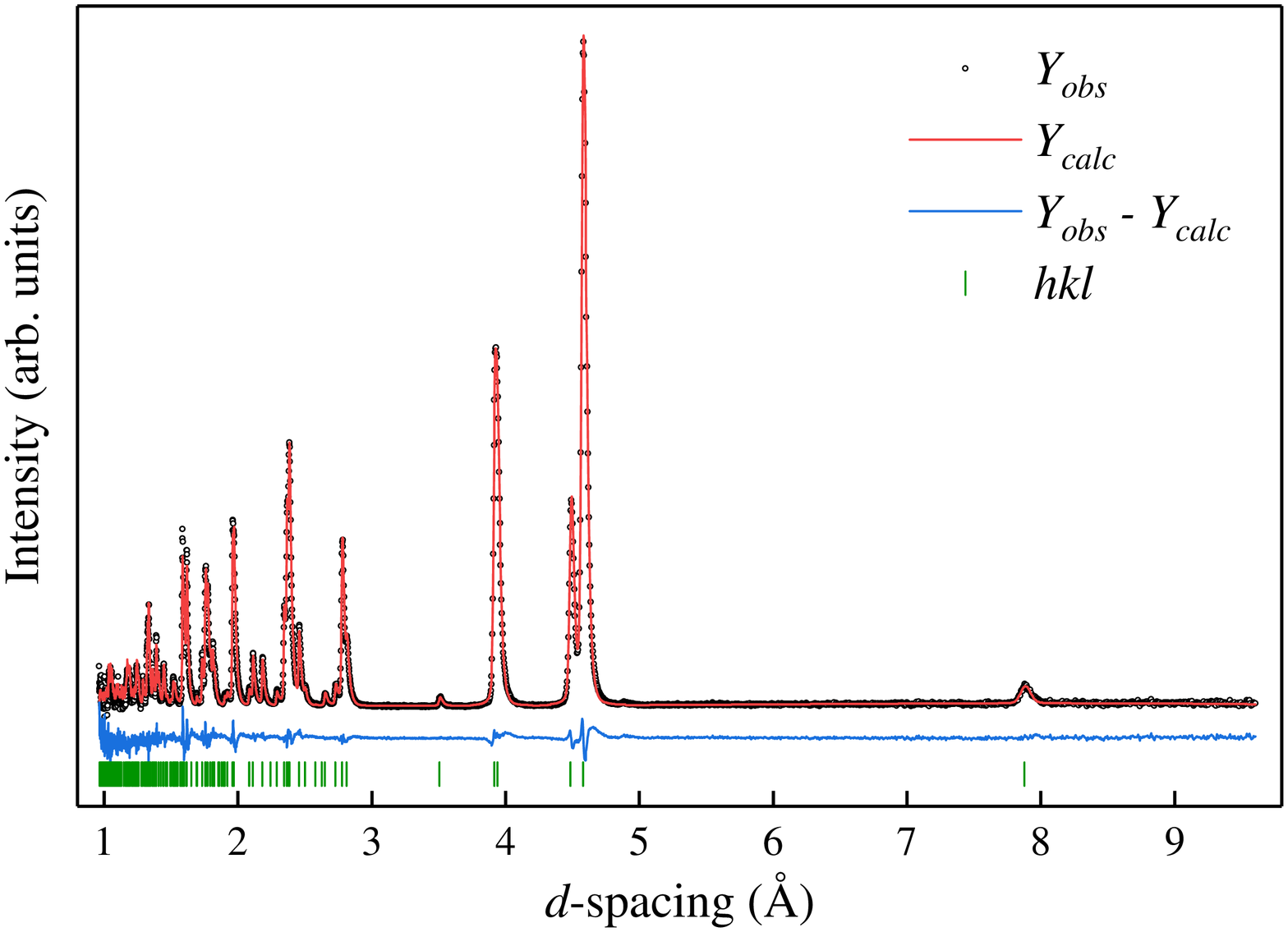}
  \caption{Measured,  calculated and  difference  curve from  Rietveld
    refinement of the magnetic structure  of $\mathrm{NaFeF_3}$ at 2 K
    for the second detector bank. The green ticks indicate reflections
    allowed by the magnetic symmetry (space group $Pn'ma'$).}
  \label{fig:NPD}
\end{figure}

\begin{figure}[ht!]
  \centering
  \includegraphics[scale=.23]{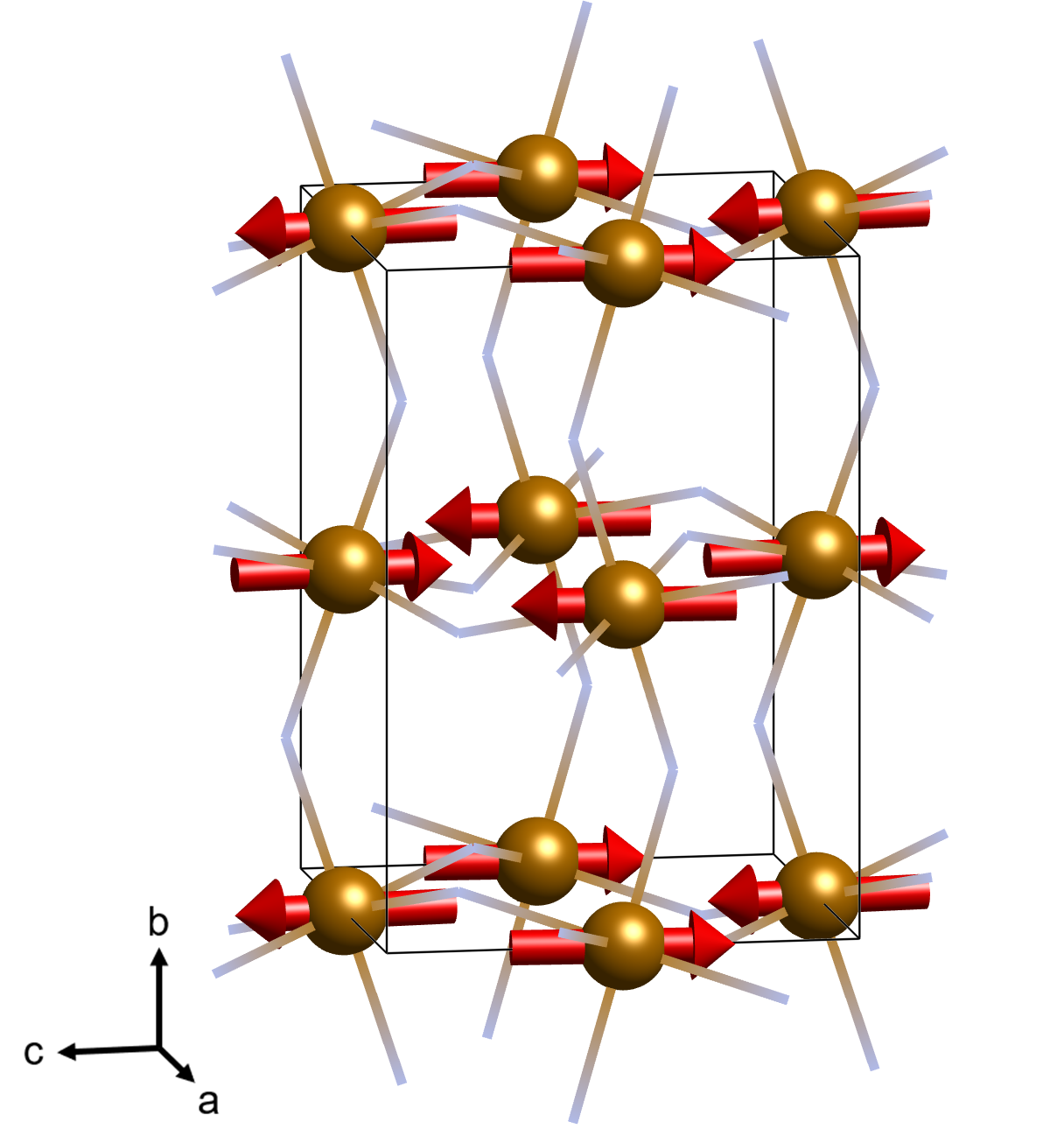}
  \caption{Magnetic structure of  the $\mathrm{NaFeF_3}$ with magnetic
    space group $Pn'ma'$.  The magnetic moments of the  iron atoms are
    antiferromagnetically ordered relative to their nearest neighbors,
    yielding G-type antiferromagnetism. Iron atoms are shown in orange
    and the  bonds to fluorine  as pale  grey lines. Sodium  atoms are
    removed for clarity.}
  \label{fig:magstr}
\end{figure}

\begin{table}[h!]
  \centering 
  \caption{Basis  functions  of   the  one-dimensional  $\Gamma$-point
    irreducible representations found by decomposition of the magnetic
    representation for  the iron  site in $\mathrm{NaFeF_3}$.  The +
    and -  symbols denote the  relative sign of the  magnetic moment
    along $x$, $y$, and $z$ on the respective site. }
  \label{tab:irrep} 
  \begin{tabular}{c c c c c c c c c c c c c c c c c c c c}
    \hline\hline
&&&&&&&&&&&&&&&&&&&\\    
& Coordinate & & &  & &  $ \Gamma^+_1 $  & & &  & $ \Gamma^+_2 $  & & & $ \Gamma^+_3 $  &  &  & & $ \Gamma^+_4 $  & & \\

$x$	 & $y$		& $z$ 		&	&  	&	$x$   		&	$y$   	&	$z$	& &	$x$   	&	$y$   	&	$z$	& &	$x$   	&	$y$   	&	$z$	& &	$x$   	&	$y$   	&	$z$   \\

0.0	 & 0.0	& 0.5	&  	&	&     +   	&	+   	&	+	& &	+   	&	+   	&	+	& &	+   	&	+   	&	+	& &	+   	&	+   	&	+   \\

0.0	 & 0.5	& 0.5	&  	&	&     -   	&	+   	&	-	& &	+   	&	-   	&	+	& &	+   	&	-   	&	+	& &	-   	&	+   	&	-   \\

0.5	 & 0.5	& 0.0 	& 	&	&     +   	&	-   	&	-	& &	-   	&	+   	&	+	& &	+   	&	-   	&	-	& &	-   	&	+   	&       +   \\

0.5	 & 0.0	& 0.0 	& 	&	&     -   	&	-   	&	+	& &	-   	&	-   	&	+	& &	+   	&	+   	&	-	& &	+   	&	+   	&	- \\
&&&&&&&&&&&&&&&&&&&\\
    \hline
  \end{tabular}
\end{table}

The magnetic  ordering of  $ \Gamma^+_4  $ can  be described  as A$_x$
antiferromagnetic ordering  along [001], F$_y$  ferromagnetic ordering
along  [010]   and  G$_z$  antiferromagnetic  ordering   along  [001],
corresponding  to  the  refined  parameters  $M_x$,  $M_y$  and  $M_z$
respectively  (\tref{irrep}).  In  our  magnetic Rietveld  refinements
(\tref{str_2K}), we find a large value for the G$_z$ component ($M_z$)
and a  small value for  the A$_x$ component ($M_x$).   When performing
Rietveld refinements of the ferromagnetic  F$_y$ component, a value of
$M_y =  0.75(3) ~ \mu_B$  is obtained for the  2 K NPD  data. However,
since  F$_y$-scattering coincides  with nuclear  peak positions,  such
quite small values of $M_y$ yield very minor changes to the calculated
diffraction  pattern and  agreement  factors.  Furthermore, these  are
strong correlations with other parameters of the refinements, e.g. the
thermal  displacement   parameter  of   Fe.   The   ZFC  magnetization
measurement furthermore suggests that  the ferromagnetic moment should
almost  absent at  low temperatures  without  a magnetic  field. As  a
consequence, the NPD  data cannot be used to claim  the existence of a
F$_y$  component. Hence,  $M_y$ was  fixed  to zero  during the  final
refinements.

 \begin{table*}[t]
  \centering
  \caption{Atomic  coordinates  of  $\mathrm{NaFeF_3}$  from  Rietveld
    refinement of NPD at 2 K. The refinement was performed in magnetic
    space group $Pn'ma'$ with lattice  parameters of $a$ = 5.62571(11)
    \AA, $b$ = 7.87673(15) \AA and $c$ = 5.45623(10) \AA, and magnetic
    parameters $M_x$ = 0.422(11) $\mu_B$,  $M_y$ = 0 $\mu_B$ and $M_z$
    = 4.221(4)  $\mu_B$, yielding a  total ordered magnetic  moment of
    $M$ = $4.246(11)~\mu_B$.  }
  \label{tab:str_2K}
  \begin{tabular}{l c c c c c l}
         &&&&&&\\
    \hline\hline
         &&&&&&\\
    Site & Multiplicity & $x$ & $y$ & $z$ & Occ & U$_{iso}$ (\AA$^2$) \\
         &&&&&&\\
    Na &  4& 0.0544(2) 		& 0.25 			& 0.9826(3) 	& 1 & 0.0175(4)           \\
    Fe &  4& 0.5 			& 0 			& 0 			& 1 & 0.00438(17)     \\
    F1 &  4& 0.45061(16) 	& 0.25 			& 0.11401(16) 	& 1 & 0.0105(3)             \\
    F2 &  8& 0.29707(11) 	& 0.06114(8) 	& 0.68918(11) 	& 1 & 0.0096(2)         \\
         &&&&&&\\
    \hline
  \end{tabular}
\end{table*}

Complete structural details obtained  from the Rietveld refinements of
NPD  data is  given in  \tref{str_2K} and  \tref{moment}. The  refined
model  has a  dominating G$_z$-type  antiferromagnetic structure  with
moments aligned  parallel to [001].  There are weak indications  for a
small  canting parallel  to  [100]  as given  by  the A$_x$  component
(\fref{magstr}).   The  magnetic   moments  are  antiferromagnetically
oriented   with    respect   to   their   nearest    neighbors   along
$\mathrm{[010]}$,                 $\mathrm{[101]}$                 and
$\mathrm{[\bar{1}01]}$.  Neighboring spins  are aligned  close to  the
equatorial  plane of  the octahedra  (defined  by the  four F2  atoms,
\fref{str}) in the crystallographic (010)-plane.

At 2  K the Rietveld  refinements give  an ordered magnetic  moment of
$4.246(11)~\mu_B$, slightly lower that the theoretical spin-only value
for  $\mathrm{Fe^{2+}}$  of  $4.90~\mu_B$.    The  value  is  in  good
agreement with that of  $\mathrm{Fe^{2+}}$ in $\mathrm{KFeF_3}$, which
is  $4.42~\mu_B$  \cite{GTYPE}.  The  slightly  lower  value than  the
spin-only value can be accounted for by a slight hybridization between
iron  and fluorine,  effectively reducing  the number  of magnetically
ordered  electrons.  A  very weak  additional ferromagnetic  component
will not change this picture, but should not be neglected, see below.

The refinements (and magnetic peak  intensities) show that the ordered
magnetic moment decreases upon heating  from 2~K. This is evidenced in
the refined values for $M_x$ and $M_z$ (\fref{str_prm}a). The magnetic
reflection in NPD disappears at 95  K, which is in compliance with the
N\'eel temperature found by magnetometry.

We  note  the  presence  of a  strong  magnetostructural  coupling  in
$\mathrm{NaFeF_3}$,  evidenced   by  major  changes  in   the  lattice
parameters at the N\'eel ordering temperature (\fref{str_prm}b and c).
At the transition, we observe a  rather smooth contraction of the unit
cell  volume, however,  this is  an average  of a  contraction of  the
$a$-axis in contrast to expansion of the $b$- and $c$-axis.

\begin{figure}[t!]
  \centering
  \includegraphics[scale=.29]{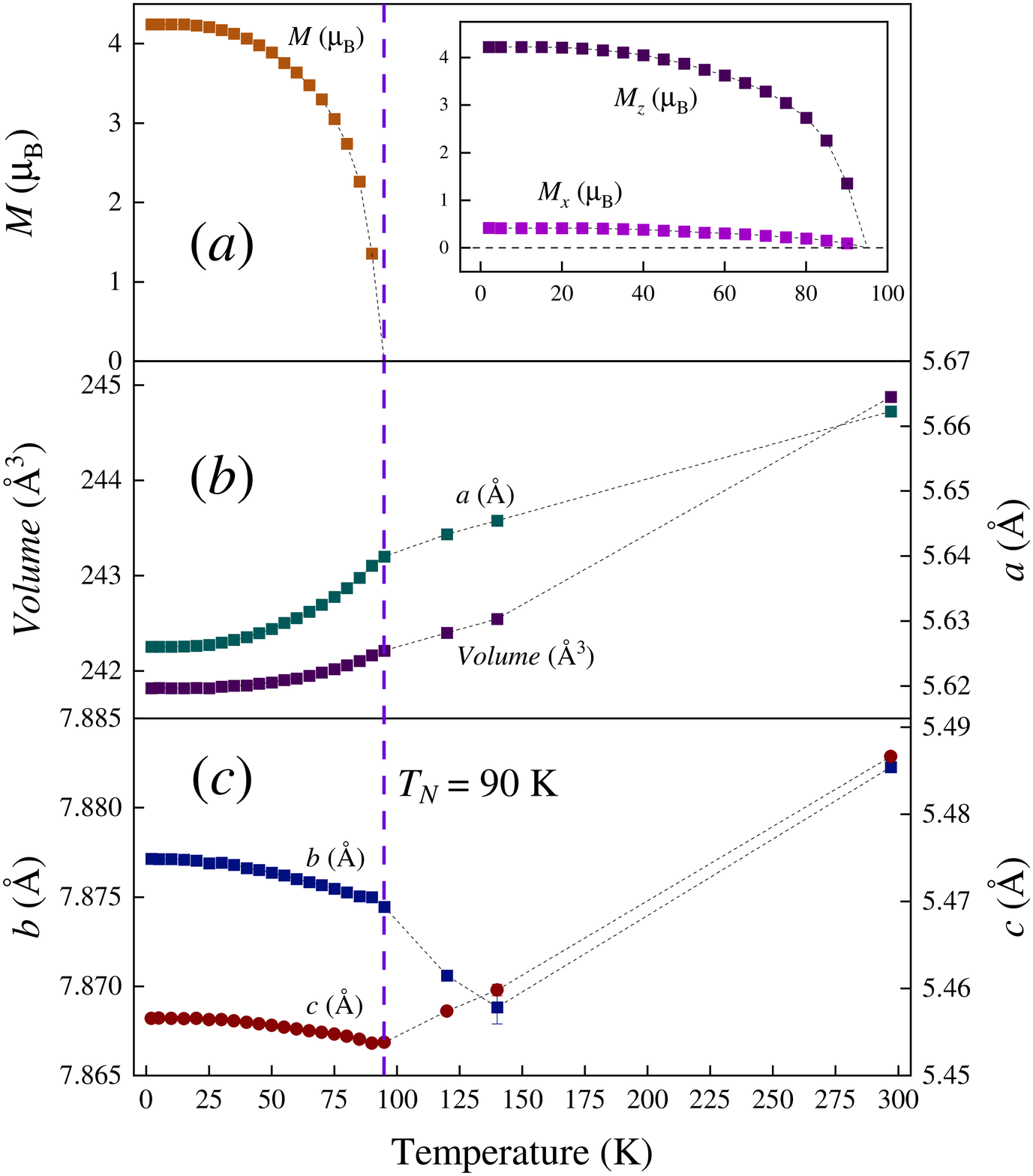}
  \caption{  Temperature dependence  of structural  parameters derived
    from Rietveld refinements  of NPD; (a) the  total magnetic moment,
    the  inset  show  the  $M_x$ and  $M_z$  components,  (b)  lattice
    parameter $a$ and  cell volume, and (c) lattice  parameter $b$ and
    $c$. The  N\'eel temperature of  90 K  is indicated by  the purple
    dashed line.  }
  \label{fig:str_prm}
\end{figure}

Already at  120 K, well above  the ordering temperature, we  observe a
change     in    temperature     dependence    for     the    $b$-axis
(\fref{str_prm}).  Actually, by  evaluating  the difference  intensity
plot between  NPD patterns at  95 and 120~K  (close to, and  above the
magnetic ordering  temperature respectively), we observe  a broad peak
at around 4.8 \AA~(\fref{NPD_diff}), which corresponds to the position
of  the (011)  and (110)  magnetic  Bragg reflections  in the  ordered
phase.  The broad peak at 4.8  \AA~ is thus interpreted as originating
from the  existence of  short-range magnetic  ordering just  above the
N\'eel   temperature.   ${\rm   Fe^{2+}}$  is   reported  to   display
magnetostrictive  behavior  and  diffuse  scattering  above  $T_N$  in
$\mathrm{Rb_2FeF_4}$,  and we  note  that the  diffuse scattering  for
$\mathrm{NaFeF_3}$ above $T_N$ also coincides with a tensile effect on
the lattice \cite{MAGSTRICRP1}.

\begin{figure}[t!]
  \centering
  \includegraphics[scale=.34]{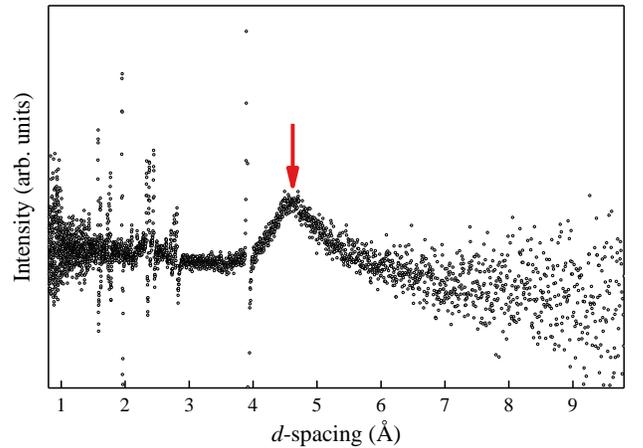}
  \caption{Plot of the difference between NPD patterns at 95 and 120 K
    measured on the low resolution  detector bank. A broad peak around
    4.8 \AA~ is marked by a red arrow, which corresponds to the (011)-
    and  (110)-magnetic Bragg  reflections  and indicates  short-range
    magnetic order appearing at 95 K.}
  \label{fig:NPD_diff}
\end{figure}

As   discussed  above,   due   to  a   weak  Jahn-Teller   distortion,
$\mathrm{NaFeF_3}$  adopts   short,  medium   and  long   Fe-F  bonds,
\fref{str}. Considering the variation of  the Fe-F bond length between
2 and 297 K in the NPD  experiment, we observe a clear trend: The bond
lengths  tend  towards  receiving identical  values  when  temperature
approaches  297   K  (\fref{NPD_bond}).    In  the   cubic  perovskite
$\mathrm{KFeF_3}$, the Fe-F  bond lengths adopts a value  of 2.06 \AA,
which ought to be  a value expected also for the  Fe-F bond lengths in
$\mathrm{NaFeF_3}$  \cite{KFEF3}.  On this  basis  it  is tempting  to
suggest  that $\mathrm{NaFeF_3}$  may  undergo  transitions to  higher
symmetric structures at elevated temperatures \cite{ME1,ME2}.  This is
beyond the scope of the present work.

\begin{figure}[t!]
  \centering
  \includegraphics[scale=.30]{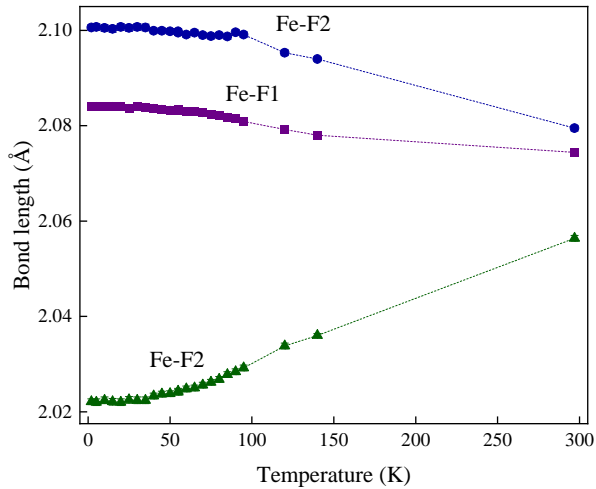}
  \caption{Variation of  the Fe-F  bond lengths derived  from Rietveld
    refinements of NPD between 2  and 297 K. Uncertainties are smaller
    than the symbol size. }
  \label{fig:NPD_bond}
\end{figure}

\section{Discussion}
\label{sec:DIS}

We highlight that our new  wet chemical synthesis protocol has allowed
preparation  of $\mathrm{NaFeF_3}$  of  very high  purity  which is  a
prerequisite for  clarification of intrinsic  properties. Importantly,
we compared  ZFC - FC  magnetic behaviour  of phase pure  samples with
reported data  for samples made  by a conventional  solid-state method
\cite{ME1}.  For  $\mathrm{NaFeF_3}$ prepared by  solid-state methods,
iron  impurities   easily  outweigh  the  weak   antiferromagnetic  or
paramagnetic signal, resulting  in overestimated magnetization values.
This  is mirrored  by  a  measured ferromagnetic  behaviour  at 300  K
\cite{ME1}, which is not intrinsic to pure $\mathrm{NaFeF_3}$ prepared
by our wet-chemical  method.  Note, that the  very weak ferromagnetism
observed  in AC  susceptibility below  the N\'eel  temperature of  our
phase  pure $\mathrm{NaFeF_3}$  samples is  intrinsic, but  with value
close to zero.

Although NPD  cannot unambiguously conclude on  the possible existence
of  a symmetry  allowed  ferromagnetic component  along [010]  (F$_y$)
according  to  the magnetic  space  group  $Pn'ma'$, this  is  clearly
indicated by a strong peak in the imaginary component $\chi ''$ of the
AC susceptibility. Considering the magnetic F - Fe - F interactions of
$\mathrm{NaFeF_3}$ in light  of the Goodenough-Kanamori-Anderson (GKA)
rules,  antiferromagnetism  is  expected  \cite{GKA}.  Furthermore,  a
linear   G-type  magnetic   structure   was  predicted   by  DFT   for
$\mathrm{NaFeF_3}$ \cite{ME1}, also with  moments arranged along [001]
(G$_z$-type). However, our Rietveld  analysis considered an additional
weak    A$_x$-component,     which    implies    canting     of    the
antiferromagnetism.  The latter  component is  close to  the detection
limit of the analysis.

G-type   magnetic    ordering   is    observed   in    several   other
fluoroperovskites, e.g.   $\mathrm{KMF_3}$ (M  = Mn,  Fe, Co,  and Ni)
\cite{GTYPE}. However, due  to the small sodium cation  and the weakly
Jahn-Teller    active    $\mathrm{Fe^{2+}}$,    the    structure    of
$\mathrm{NaFeF_3}$  is significantly  distorted  and  the F-Fe-F  bond
angles deviate  from 180$^\circ$.  As a  consequence, the interactions
may deviate from the GKA rules.

Correspondingly,  one must  consider  other  magnetic interactions  as
origin to the  weak ferromagnetism. For compounds  with $d^6$ electron
configuration,   Jahn-Teller,  as   well  as   spin-orbit  interaction
mechanism, will  contribute to stabilization of  the system \cite{KK}.
If   spin-orbit    coupling   is   present    in   $\mathrm{NaFeF_3}$,
Dzyaloshinskii-Moriya interactions  may occur. Such  interactions give
rise to  ferromagnetic exchange  and may  thus be  the origin  of weak
ferromagnetism in $\mathrm{NaFeF_3}$ \cite{DINTER,MINTER}.



\section{Conclusion}
\label{sec:CON}

In summary  we have developed  a wet-chemical synthesis  protocol that
allow  preparation of  $\mathrm{NaFeF_3}$ in  large quantities  and of
high purity.   As a consequence we  have been able to  investigate the
intrinsic magnetic properties  of $\mathrm{NaFeF_3}$ without potential
additional    magnetic    contributions     from    impurities    like
$\mathrm{\alpha}$-Fe that  will interfere with the  analysis. Magnetic
susceptibility  and powder  neutron  diffraction  analysis shows  that
$\mathrm{NaFeF_3}$ has a  N\'eel temperature of 90  K. AC magnetometry
clearly show  the presence of  weak ferromagnetism below  the ordering
temperature,  supported by  field dependent  DC measurements.  Neutron
diffraction data describe  the compound as a  weakly canted G$_z$-type
antiferromagnet with a minor  A$_x$-component allowed by symmetry. The
magnetic space  group opens  for a F$_y$  component, however,  this is
almost absent at  zero-field and is too weak to  be proven the current
analysis.  The  temperature variation  of  the  Fe-F bonds  suggest  a
possible structural  phase transition to a  higher symmetric structure
above 300 K.

\section{Aknowledgements}
\label{sec:AK}

We  thank  Serena Margadonna  (Swansea  University,  Swansea, UK)  for
providing project support  via the Research Council  of Norway project
214260. This  work was partially performed  within the RIDSEM-project,
financed by the Research Council  of Norway (Project No. 272253).  The
U.K. Science and  Technology Facilities Council (STFC)  is thanked for
allocating beamtime at the ISIS Facility.  We also thank Pascal Manuel
for help during the NPD experiment, and Asbj\o rn Slagtern Fjellv\aa g
and Vincent Hardy for discussions regarding magnetic properties.

\bibliographystyle{plain}
\bibliography{NaFeF3_arXiv}

\begin{thebibliography}{}

\bibitem{FLUOR}
  M.~Leblanc, V.~Maisonneuve and A.~Tressaud
\newblock{\em Chem. Rev.}, {\bf 115}, 1191--1254, 2015

\bibitem{FLUOR2}
A.~Tressaud
\newblock{\em J. Fluorine Chem.}, {\bf 132}, 1191--1254, 2011

\bibitem{MultiF1}
R.M. Dubrovin, L.N. Alyabyeva, N.V. Siverin, B.P. Gorshunov, N.N. Novikova, K.N. Boldyrev, and R.V. Pisarev
\newblock{\em Phys. Rev. B}, {\bf 101}, 180403, 2020

\bibitem{MultiF2}
R.M. Dubrovin, S.A. Kizhaev, P.P. Syrnikov, J.-Y. Gesland, and R.V. Pisarev
\newblock{\em Phys. Rev. B}, {\bf 98}, 060403, 2018

\bibitem{SPINTRON1}
  H.~B\'ea, M.~Gajek, M.~Bibes and A.~Barth\'el\'emy
\newblock{\em J.  Phys. Condens. Mat.}, {\bf 20}, 434221, 2008

\bibitem{SPINTRON2}
I.~\v{Z}uti\' c, J.~Fabian and S.D.~Sarma
\newblock{\em Rev. Mod. Phys.  }, {\bf 76}, 323  ,  2004

\bibitem{NF3SIB}
A.~Kitajou,  H.~Komatsu, K.~Chihara,  I.D.~Gocheva,  S.~Okada and   J.~Yamaki
\newblock{\em J. Power Sources}, {\bf 198}, 389--392, 2012.

\bibitem{NF3SIB2}
Y.~Yamada, T.~Doi, I.~Tanaka,  S.~Okada and   J.~Yamaki
\newblock{\em J. Power Sources}, {\bf 196}, 4837--4841, 2011.

\bibitem{NCATH}
K.V.~Kravchyk, T.~Z\u nd, M.~Worle, M.V.~Kovalenko and M.I.~Bodnarchuk
\newblock{\em Chem. Mater}, {\bf 30}, 1825--1829, 2018.

\bibitem{ME1}
F.L.M.~Bernal, K.V.~Yusenko, J.~Sottmann, C.~Drathen, J.~Guignard, O.M.~l\o vvik, W.A.~Chrichton and S.~Margadonna
\newblock {\em Inorg. Chem}, {\bf 53}, 12205--12214, 2014.

\bibitem{ME2}
W.A.~Chrichton, F.L.M.~Bernal, J.~Guignard, M.~Hanfland  and S.~Margadonna
\newblock {\em Mineralogical Magazine}, {\bf 80}, 659--674, 2016.

\bibitem{PPV}
  J.M.~De Teresa, M.R.~Ibarra, P.~Algarabel, L.~Morellon, B.~García-Landa, C.~Marquina, 
  C.~Ritter, A.~Maignan, C.~Martin, B.~Raveau, A.~Kurbakov, and V.~Trounov
\newblock {\em Phys. Rev. B}, {\bf 65}, 100403, 2002.

\bibitem{WISH}
L.C.~Chapon, P.~Manuel.
\newblock{\em Neutron News}, {\bf 22}, 22, 2011

\bibitem{MANTID} O.  Arnold, J.C.  Bilheux, J.M.  Borreguero,
  A. Buts, S.I. Campbell, L.  Chapon, M. Doucet, N. Draper, R. Ferraz  Leal, M.A. Gigg, V.E.  Lynch, A. Markvardsen, D.J.  Mikkelson, R.L. Mikkelson, R. Miller,  K. Palmen, P. Parker, G. Passos, T.G. Perring, P.F. Peterson,
  S.  Ren, M.A.  Reuter, A.T.  Savici,  J.W.  Taylor,
  R.J. Taylor, R. Tolchenov, W. Zhou and J. Zikovsky
\newblock{\em Nucl. Instrum. Methods Phys. Res. A}, {\bf 764}, 156, 2014.

\bibitem{JANA}
  V.~Pet\v{r}\'i\v{c}ek, M.~Du\v{s}eka and L.~Palatinus
\newblock{\em Z.Kristallogr. Cryst. Mater}, {\bf 229}, 345, 2014  

\bibitem{ME3}
  F.L.M.~Bernal, J.~Sottmann, D.S.~Wragg, H.~Fjellv\aa g, \O .S.~Fjellv\aa g, C.~Drathen, W.A.~S\l awi\'{n}ski, and O.M.~L\o vvik
\newblock{\em Phys. Rev. Materials}, {\bf 4}, 054412 , 2020

\bibitem{ACmag}
  A.~Szytu\l a, M.~Ba\l anda, B.~Penc and M.~Hofmann
\newblock{\em J. Phys.: Condens. Matter}, {\bf 12}, 7455--7462, 2000



\bibitem{MAGSTRICRP1}
R.J.~Birgeneau, H.J.~Guggenheim and G.~Shirane
\newblock{\em 	Phys. Rev. B }, {\bf 1}, 2211, 1970.

\bibitem{KFEF3}
R.J.~Birgeneau, H.J.~Guggenheim and G.~Shirane
\newblock{\em 	Acta Cryst. B }, {\bf 39}, 561--564, 1983.

\bibitem{GKA}
 J.B.~Goodenough
 \newblock{\em Magnetism and the Chemical Bond}, R. E. Krieger Publishing Company, 1976

\bibitem{GTYPE}
  V.~Scatturin, L.~Corliss, N.~Elliott and J.~Hastings
\newblock{\em Acta Cryst.}, {\bf 14}, 19--26, 1961


\bibitem{KK}
  K.I.~Kugel' and D.I.~Khomski\u{\i}
\newblock{\em Sov. Phys. Usp.}, {\bf 25}, 231, 1982


\bibitem{DINTER}
  I.~Dzyaloshinsky
\newblock{\em 	J. Phys. Chem. Solids}, {\bf 4}, 241--255, 1958

\bibitem{MINTER}
  T.~Moriya
\newblock{\em Phys. Rev.}, {\bf 120}, 91, 1960







\end{thebibliography}

\newpage

\begin{figure*}[t!]
  \centering
  \includegraphics[scale=.34]{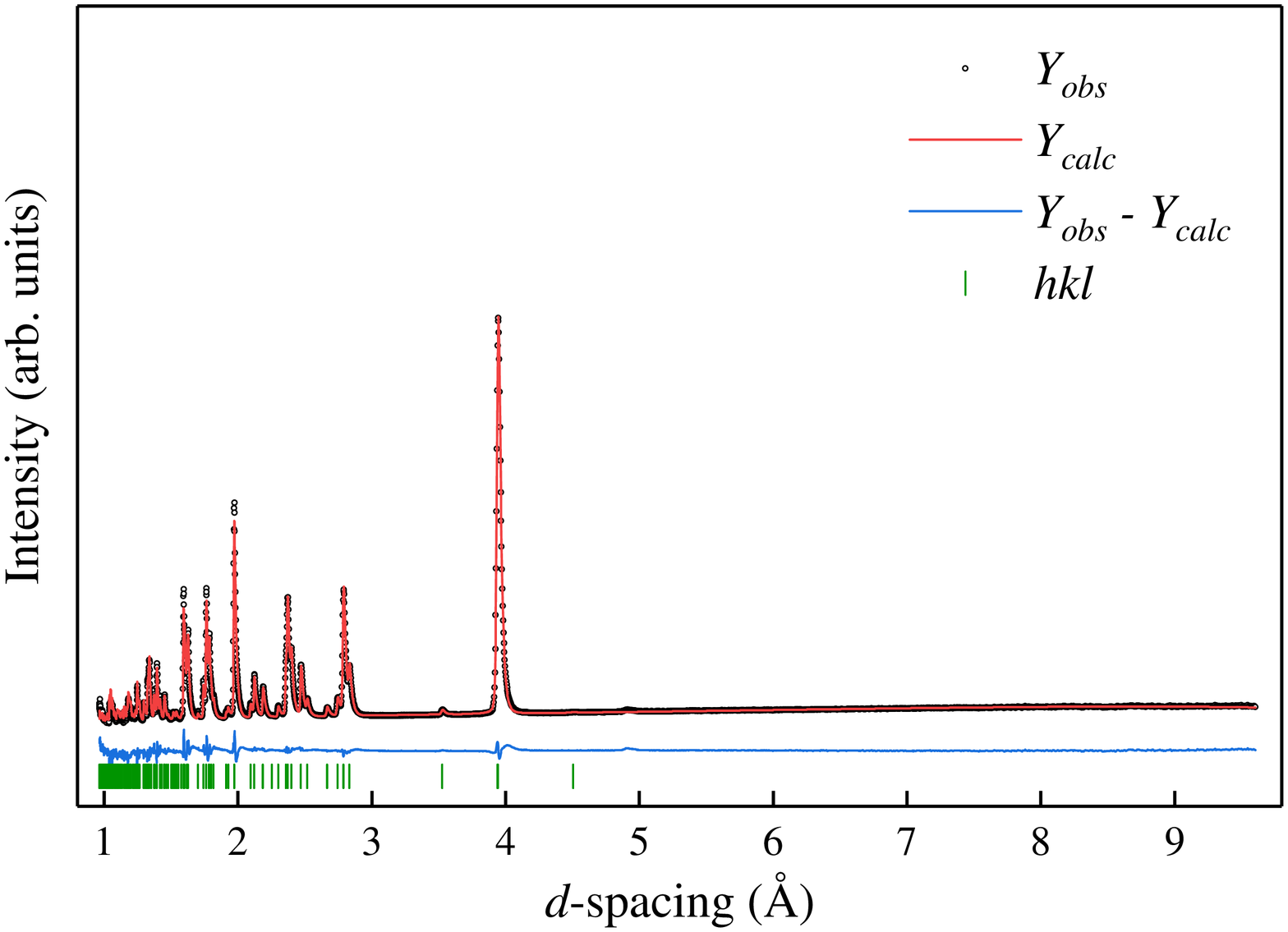}
  \caption{Measured,  calculated and  difference  curve from  Rietveld
    refinement of the nuclear structure of $\mathrm{NaFeF_3}$ at 297 K
    for the second detector bank.  The green tics indicate reflections
    allowed by the space group ($Pnma$).}
  \label{fig:NPD_RT}
\end{figure*}

\begin{table*}[h!]
  \centering 
  \caption{Atomic coordinates  and magnetic  moments $M_x$,  $M_y$ and
    $M_z$  for   Fe  in   $\mathrm{NaFeF_3}$  derived   from  Rietveld
    refinements of NPD data below  $T_N$. The refinement was performed
    in magnetic space group $Pn'ma'$.  }
  \label{tab:moment} 
  \begin{tabular}{c c c c c c c c c c c c c c c c c c c c c c}
    \hline\hline
    &&&&\\
Temperature (K) &&&& $x$ && $y$ && $z$ &&&& $M_x$ &&& $M_y$ &&& $M_z$ \\
\hline \\
2	&&&&  0.5 &&  0   &&  0   &&&&	0.42(1)	&&& 0 &&&	4.224(4)	\\
5	&&&&  0.5 &&  0   &&  0   &&&&	0.41(1)	&&& 0 &&&	4.223(4)	\\
10	&&&&  0.5 &&  0   &&  0   &&&&	0.41(1)	&&& 0 &&&	4.223(4)	\\
15	&&&&  0.5 &&  0   &&  0   &&&&	0.41(1)	&&& 0 &&&	4.223(4)	\\
20	&&&&  0.5 &&  0   &&  0   &&&&	0.41(1)	&&& 0 &&&	4.210(4)	\\
25	&&&&  0.5 &&  0   &&  0   &&&&	0.41(1)	&&& 0 &&&	4.189(4)	\\
30	&&&&  0.5 &&  0   &&  0   &&&&	0.40(1)	&&& 0 &&&	4.153(4)	\\
35	&&&&  0.5 &&  0   &&  0   &&&&	0.39(1)	&&& 0 &&&	4.108(4)	\\
40	&&&&  0.5 &&  0   &&  0   &&&&	0.38(1)	&&& 0 &&&	4.048(4)	\\
45	&&&&  0.5 &&  0   &&  0   &&&&	0.36(1)	&&& 0 &&&	3.964(4)	\\
50	&&&&  0.5 &&  0   &&  0   &&&&	0.34(1)	&&& 0 &&&	3.875(4)	\\
55	&&&&  0.5 &&  0   &&  0   &&&&	0.32(1)	&&& 0 &&&	3.743(3)	\\
60	&&&&  0.5 &&  0   &&  0   &&&&	0.30(1)	&&& 0 &&&	3.624(3)	\\
65	&&&&  0.5 &&  0   &&  0   &&&&	0.28(1)	&&& 0 &&&	3.468(3)	\\
70	&&&&  0.5 &&  0   &&  0   &&&&	0.25(1)	&&& 0 &&&	3.288(3)	\\
75	&&&&  0.5 &&  0   &&  0   &&&&	0.22(1)	&&& 0 &&&	3.045(3)	\\
80	&&&&  0.5 &&  0   &&  0   &&&&	0.19(1)	&&& 0 &&&	2.733(3)	\\
85	&&&&  0.5 &&  0   &&  0   &&&&	0.15(2)	&&& 0 &&&	2.256(3)	\\
90	&&&&  0.5 &&  0   &&  0   &&&&	0.09(4)	&&& 0 &&&	1.354(4)	\\
    \hline
  \end{tabular}
\end{table*}

\end{document}